\documentclass[10pt, conference, letterpaper]{IEEEtran}
\usepackage{graphicx}
\usepackage{caption}
\usepackage{subcaption}
\usepackage{color}
\usepackage{amsmath}
\usepackage{algorithm}
\usepackage{algpseudocode}

\usepackage{url}
\usepackage{bm}

\begin{document}

\title{Exploiting Network Awareness to Enhance DASH Over Wireless}

\author{
\IEEEauthorblockN{Francesco Bronzino$^{\S}$, Dragoslav Stojadinovic$^{\S}$, Cedric Westphal$^{\dag}$, Dipankar Raychaudhuri$^{\S}$}\\
\IEEEauthorblockA{
	$^{\S}$WINLAB, ECE Department, Rutgers University, New Brunswick, NJ, USA\\
	$^{\dag}$Innovation Center, Huawei Technology, Santa Clara, CA, USA\\
	$^{\dag}$Computer Engineering Deptartment, University of California, Santa Cruz, CA, USA\\
	Email: $^{\S}$\{bronzino,stojadin,ray\}@winlab.rutgers.edu, $^{\dag}$cwestphal@huawei.com}
}

\maketitle

\begin{abstract}

The introduction of Dynamic Adaptive Streaming over HTTP (DASH) helped reduce the consumption of resource in video delivery, but its client-based rate adaptation is unable to optimally use the available end-to-end network bandwidth. We consider the problem of optimizing the delivery of video content to mobile clients while meeting the constraints imposed by the available network resources. Observing the bandwidth available in the network's two main components, core network, transferring the video from the servers to edge nodes close to the client, and the edge network, which is in charge of transferring the content to the user, via wireless links, we aim to find an optimal solution by exploiting the predictability of future user requests of sequential video segments, as well as the knowledge of available infrastructural resources at the core and edge wireless networks in a given future time window. Instead of regarding the bottleneck of the end-to-end connection as our throughput, we distribute the traffic load over time and use intermediate nodes between the server and the client for buffering video content to achieve higher throughput, and ultimately significantly improve the Quality of Experience for the end user in comparison with current solutions.

\end{abstract}

\section{Introduction}


Video consumption over the internet has experienced a continuous growth in the last few years and now accounts for about two thirds of the total amount of global Internet traffic. Its share is expected to increase to up to 79\% by 2018 \cite{ciscoForecast}. While video streaming services continue to raise their popularity thanks to large availability of content and reduced costs, Internet Service Providers are struggling to provide high quality services to their costumers due to their inability to allocate enough capacity to meet such demand, especially at peak hours.

To adjust the demand to the network conditions, Dynamic Adaptive Streaming over HTTP (DASH) has been introduced. DASH let the client adjust the rate of the video stream by monitoring the network conditions perceived by this stream. DASH is a client-based rate adaptation, as the server is stateless and the network is considered as a black box. The client measures the end-to-end throughput between itself and the server and modifies the rate accordingly.

While this approach has been incredibly successful and DASH-like rate adaptation now accounts for most of the video traffic (say, Netflix or YouTube), end-to-end rate adaptation is suboptimal. Indeed, end-to-end monitoring measures the lowest throughput of all the links in between the client and the server, while the available bandwidth on these links varies a lot. 

\begin{figure}[!t]
\centering
\includegraphics[width=\linewidth]{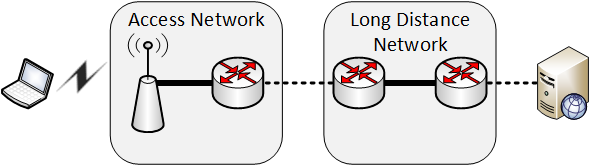}
\caption{Basic network representation.}
\label{fig:base_network}
\vspace{-4mm}
\end{figure}

As a simple illustrative example, consider the following scenario of Figure~\ref{fig:base_network}: the client is connected to the server by two links (say, a wireless link for the edge network, and a wired link to the server). Both of the links' bandwidth will oscillate and in modern networks, both could be congested\footnote{It is well known for instance that Netflix bandwidth is throttled inside the network by operators with whom Netflix does not have a peering agreement.}. Therefore an end-to-end mechanism will yield the minimal available bandwidth of each link. If for one unit of time, the capacity of the wireless is 1 unit of transfer, while the capacity of the wired link is 2, and for the next unit of time the capacities are reversed, then the end-to-end mechanism can only achieve a rate of 1. 

However, inserting an intermediary node with storage capacity in between the two links could increase the throughput significantly. In the illustrative example, during the first time slot, the full capacity of the wired link could be used to delivery 1 unit to the client and 1 unit to the intermediate storage; in the second time slot, this extra stored unit can be delivered over the air interface, for a total delivery of 2 units. On average, the delivery is of 1.5 for each time slot, versus 1 in the end-to-end mechanism. While this example is simplistic, and does not take into account rate adaptation, it shows the benefit of a network-assisted content deliver mechanism. Here we present an intermediated rate adaptation mechanism which leverages the different time-varying capacities of the multiple links in the network.

The goal of our work is to propose a novel infrastructural approach to control the load on the network for clients with quickly changing available capacity while maintaining high quality experience for the end users. The experience is defined as a set of quality metrics, such as average bitrate, temporal variability and amount of rebuffering time. Our solution is based on two key ideas: (1) to use local caches strategically deployed into the access networks to decouple the transmission speed of the Internet long distance network from the local access network and (2)to exploit the predictability of future available infrastructural resources (at the core and edge wireless network) and the predictability of video requests to distribute the load on the network of the video streams and control the Quality of Experience for the end users.

The rest of paper is structured as follow: in Section \ref{sec:related} we will discuss our contributions in comparison to other related works; in Section \ref{sec:networkModel} we will present the network model that we use to then design our solution; Section \ref{sec:experiments} will provide a detailed experimental analysis of our design through extensive simulations and finally Section \ref{sec:conclusions} will conclude the paper.

\section{Related work}\label{sec:related}

Bianchi~\cite{bianchi1997role} has considered the role of intermediate storage in the content delivery, however in the context of ATM and without rate adaptation. They did observe a similar benefit in their application scenario. Since this work,  Dynamic Adaptive Streaming over HTTP (DASH) \cite{sodagar2011mpeg} has been introduced as a standard for video streaming. It has been widely deployed by many major video streaming services such as Netflix, Hulu, and Youtube.  

Letting the network assist in the delivery of the content is related to the ideas of the Information-Centric Network (ICN) architectures\cite{jacobson2007content}\cite{dannewitz2009netinf}\cite{koponen2007data} that have been proposed recently to allow the network to be aware of content semantics.  Some have proposed to modify DASH over ICN, for instance in the IRTF~\cite{ledereradaptive}. \cite{grandl2013interaction} examines the interaction of DASH with ICN, notes the potential issues, and identifies the synergies. \cite{lee2013svc} targets HTTP adaptive streaming (HAS) in Content-Centric Networking (CCN)~\cite{jacobson2007content} for Scalable Video Coding. 

\cite{malandrino2012proactive} has looked at how to predict the request from the users by looking a social interactions. Here, we use the natural predictability that video streaming offers . \cite{ariyasinghe2013distributed} describes a web prefetching module running on the CDN main node (controller) which downloads web contents that will be requested in the future to the LAN CDN surrogates. Our work is different as the time domain is much smaller and the prefetching utilizes the bandwidth more dynamically.  In addition, \cite{ariyasinghe2013distributed} does not specifically consider video contents. \cite{lu2013optimizing} tries to optimize the utilization of the wireless channel to delivery buffered video (i.e. streaming with availability of buffer at the receiving side) conjunct with the minimization of the rebuffering time using a bandwidth prediction model, but does not consider network caching or adaptive video streaming.

\cite{balachandran2013analyzing} analyzes the potential benefits of CDN augmentation strategies can offer for Internet video workloads using a dataset of 30 million VOD and live sessions. It has also been observed in \cite{plissonneau2012longitudinal}\cite{finamore2011youtube} that fractions of viewers typically watch only the first 10 minutes of video, around 4.5\% of users are serial early quitters, and 16.6\% of users consistently watch video to completion. This suggests that a user based prefetching policy should be a natural extension for our work.

\cite{li2012network} proposes a Network-Friendly DASH (NF-DASH) architecture for coordinating peer-assisted CDNs operated by ISPs and the traditional CDNs. \cite{liu2013peer} analyzes the waiting time and network utilization with service prioritization considering both on-demand fetching/caching and prefetching in a P2P-assisted video streaming system. \cite{liu2012optimizing} formulates the CDN assignment as an optimization problem targeting minimizing the cost based on the QoE constraints to CDN servers at different locations. \cite{liu2012case} shows by measurement the shortcomings of today's video delivery infrastructure and proposes a control plane in the network for video distribution for global resource optimization. 

Hybrid P2P-CDN video streaming enhancements have also been considered. That is, serving content from dedicated CDN servers using P2P technology.\cite{huang2008understanding}\cite{huang2007can} and telco-CDN (CDNs operated by telecommunication companies, enabling users to reach CDN caches that are closer) federation are two emerging strategies. Telco-CDN federation can reduce the provisioning cost by 95\%. Using P2P can lead up to 87.5\% bandwidth savings for the CDN during peak access hours.

Proxy-assisted caching and prefetching has been widely studied in the literature. Some approaches consider the quality of the connections \cite{jin2003network} and the usage of the client buffers \cite{miao1999proxy}. Approaches for transcoding proxy caching are also presented in \cite{li2005cache}\cite{qu2005cache} in which the proxy caches different versions of the content to handle the heterogeneous user requirements. Prefix caching \cite{sen1999proxy} caches only the frames at the beginning to minimize the average initial delay. Exponential segmentation \cite{park2001popularity} divides the video object such that the beginning of a video is cached as smaller segment. Lazy segmentation approach \cite{chen2005segment} determines the segment length according to the user access record at the late time. 

Proxy technologies have also been used to enhance QoE. \cite{yan2014qoe} uses information about wireless channel quality at the base station to provide QoE and fairness among clients. \cite{pu2012video} proposes WiDASH, a proxy for adaptive http streaming over wireless networks that implements a quadratic linear optimization problem to decide on the rate to use for each user/segment while giving higher priority to lower video rates.

\cite{chen2005segment} discusses  a segment-based proxy pre-fetching for streaming delivery. \cite{krishnamoorthi2013helping} evaluates the 1-ahead, n-ahead, and priority-based segment prefetching. The results show that if the bottleneck link is between client and proxy, all prefetching schemes achieve high cache hit rate after 2-3 client requesting a video. On the other hand, if the bottleneck link is between proxy and server, no prefetching helps. Our approach considers the link between the cache and the server and makes a pre-fetching decision accordingly.

\section{Network Model and Architecture}\label{sec:networkModel}

\begin{figure}[!t]
\centering
\includegraphics[width=\linewidth]{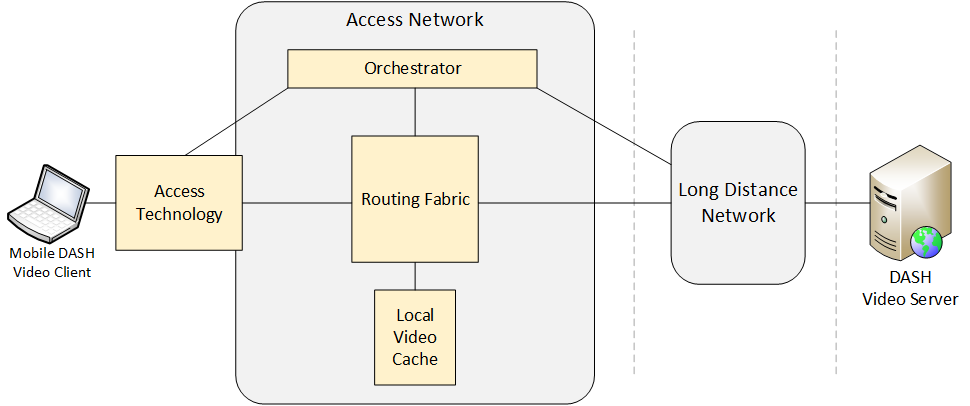}
\caption{System model.}
\label{fig:network_model}
\vspace{-5mm}
\end{figure}

We build our system around the network model shown in Figure \ref{fig:network_model} which follows the model presented in \cite{bianchi1997role}: in this classical scenario, a clients connects to the Internet through a local access network; the goal of the client is to retrieve video contents from a remote server that can be reached by traversing a long distance network. We focus on a scenario where mobile devices are connected to the access network through a wireless interface (or more than one); our system is designed to be technology agnostic so both wifi or cellular connectivity can be considered.

We base our video delivery model around the recently standardized Dynamic Adaptive Streaming over HTTP protocol. In this protocol each video file is divided into segments of equal duration encoded at different bitrates and all stored in one or more webservers. The choice of using DASH as the underlying technology for our system naturally derives from its popularity that has been constantly out-growing other protocols over the last few years thanks to its easiness of deployment as it relies on the already available HTTP infrastructure of webservers, proxies and caches. In a typical DASH system, a client interested in retrieving the video first has to retrieve from the server a Media Presentation Description file that contains information on the structure of the video, the available different bitrates for the segments and at which location they are stored. Once this step is completed, the client proceeds to sequentially download consecutive video segments. When a new segment has to be retrieved, the clients select an appropriate bitrate where the decision is usually based on different factors such as recently experience bandwidth \cite{stockhammer2011dynamic} or current buffer size \cite{huang2014buffer}.

It easy to notice that this model heavily relies on the ability of the client to estimate the available bandwidth and general network resources, a task arguably very difficult under normal network conditions \cite{huang2012confused}, and even more complex under wireless and mobile environments due to the high dynamicity caused by time-varying fading, shadowing interference and hand-off delays \cite{muller2012evaluation, riiser2012comparison}. To combat these effects we propose to move the adaptation logic where this information is more easily accessible: into the network. In our system, a caching system is made available strategically positioned inside at the edge of the architecture. An orchestration system, that from now on we will call the network controller and that could be centralized or distributed implements the adaptation logic by mean of periodically scheduling video segment requests to temporarily store the corresponding video segments into the cache; to do so, we assume that the controller, at the beginning of a periodic time window of given size, schedules the download of video segments to the available caches from where the clients will the retrieve them from. The scheduling process is based on two fundamental pieces of information to which the controller has access:
\begin{itemize}
\item A general view of the network infrastructure resources availability both in the core network and at the edge network: as different studies have proved, it is indeed possible to predict to a certain extent the variability of users connectivity in wireless networks by either exploiting movement predictability \cite{nicholson2008breadcrumbs} or by recognizing performance patterns \cite{xu2013proteus}, which is even easier if done from a convenient location inside the access network.
\item Information on the client playback and buffer status: assuming that the normal DASH client behavior is expected (i.e. sequential downloading of consecutive video segments), the controller maintains a detailed view of the current status of active video streams by tracking the progression of the requested segments; this includes the size of the video buffers available at each client and the playback time.
\end{itemize}

While this description provides a general view of our proposed architecture, we now analyze the details of each protocol and algorithm implemented in our system; we start from describing in Section \ref{sec:proto} how the basic DASH model would be modified in order to exploit the new functionalities; we then provide in Section \ref{sec:qoeModel} details on the optimization functions involved in the scheduling process and finally, Section \ref{sec:algorithms} introduces the scheduling algorithm performed by the controller.

\subsection{System Protocols Progression}
\label{sec:proto}
With the presented infrastructure we propose a solution that integrates all the given components with the goal of maximizing the Quality of Experience for the end users while only exploiting the available resources made to the system and without overrunning them. As previously introduced we use the concept of a centralized controller that orchestrates the necessary operations; while we present it as a centralized solution, we claim that the same results could be obtained with a distributed solution. Moreover, while we refer to this orchestration system as the controller, we do not limit our solution to Software Defined Networks: any technology able to track HTTP requests can be applied; possible solutions include the use of an HTTP proxy that manipulates the traffic or obviously the use of a centralized SDN controller. Our system uniquely relies on tracking and exploiting available in-network resources (i.e. capacity and caches) and tracking and eventually modifying HTTP requests from the clients. We then abstract the required actions into three main steps.

\vspace{2mm}\noindent\textbf{1) Stream initialization.} In order to initialize the streaming process, each client has to request the DASH MPD file from the server. The controller captures this requests and retrieve the same information either by deep packet inspection of the returned content or by requesting the same file from the webserver. With this information, the controller obtains a complete view of the video that can potentially be retrieved. To simplify the description and our notation, we assume that each video is composed by $N$ segments $\bm{s_i^N} = \{s_1,s_2,...,s_N\}$, each of duration $S$ seconds and available at $M$ different bitrates.

\vspace{2mm}\noindent\textbf{2) Bitrate Adaptation.} Once the MPD file is retrieved, the client can proceed to sequentially request the video segments; while in a normal DASH setup, the client would be in charge of selecting one of the available bitrates for each requested segment, we leave this duty to the controller; this design choice is justifiable by considering the fact that the controller has the best possible view of the available resources, by accessing infrastructure components and tracking the client process through its issued requests; while it would be possible to argue about the introduction of additional complexity into the network, we argue that the complexity is sustainable if the performance gains can justify it. The controller selects among the different bitrates based on two main factors: status of the video reproduction at the client (i.e. buffer size and previous quality selection) and future availability of infrastructure resources; in particular we assume that the controller has access to information regarding residual capacity available in the network and bandwidth for any specific client for a time window of size $T$ seconds.

\vspace{2mm}\noindent\textbf{3) Streaming process.} Following the normal DASH model, clients sequentially retrieve video segments by mean of issuing HTTP requests. We differentiate from the original model by having the clients to retrieve the video from the caching system without the need of specifying any particular version of the segment. This could be done using commonly exploited techniques as for example: having the network controller to modify the MPD originally retrieved replacing content URLs with locally resolvable ones or by transparently capture the HTTP requests (e.g. if the controller is implemented as an HTTP proxy) or 3) In order to meet the proposed goals the controller divides the delivery path into two components: from the server to the local caches available at the edge network and from the caches to the client over the wireless link. The controller uses the available information in the time slot of size $T$ seconds to provide in time delivery to the caches. 

Additional details regarding the involved scheduling algorithm will be provide in section \ref{sec:algorithms}, but in order to do so, we first need to introduce the Quality of Experience model employed.

\subsection{Optimization Function}\label{sec:qoeModel}

\subsubsection{QoE Model}
We include in our model the following typical key factors that play an active role in defining the QoE for video streaming:

\vspace{2mm}\noindent\textbf{Average quality}: multiple works have shown how the average video quality can not be used as the sole metric in determining the quality experienced by a user. Nevertheless, we still need to have a way of factoring the proportional quality between different available bitrates; for this reason we refer to previous work \cite{balachandran2012quest} where has been widely proven that there is a direct relationship between bitrate selection and quality governed by logarithmic laws.

\vspace{2mm}\noindent\textbf{Temporal quality variance}: different studies have proven how representation switching can factor negatively against the quality of experience; in particular, \cite{balachandran2012quest} found that only up to 0.5 quality switches per minute are considerable tolerable by users, causing exponential increase on rate of abandonments if these criteria are not met. Moreover, as suggested in the ITU standard \cite{ITUDoc}, human memory effects can distort quality ratings if noticeable impairments occur in approximately the last 10-15s of the sequence, exponentially decaying afterward, causing past factors to relatively influence the current visualized video.

\vspace{2mm}\noindent\textbf{Buffering time and ratio}: it has been widely demonstrated that frequency and length of video rebuffering highly affects the perceived quality of a video stream where each event increase the rate of abandonment and reduce the probability of client return.

\vspace{2mm}While the third factor negatively impacts the abandonment rates from users, we will work with the initial assumption that these events are occurring very infrequently; namely we assume the available bandwidth always provide at least enough resources to deliver the lowest quality video; a rebuffering cost could then be introduced into our optimization function to factor it into the scheduling algorithm.

We initially focus our attention on the first two points in defining a model of the quality of experience perceived by a client over a video session. First we we express the quality of a video segment following the logarithmic law:
$$q(r_i, r) =\alpha ln \frac{\beta r_i}{r}$$
where $r$ is the minimum quality available for the segment, $r_i$ is the quality of the considered segment and $\alpha$ and $\beta$ are specific factors that vary depending on the type of the displayed content.

Taking into consideration the second factor, we define a temporal quality variance penalty for two following video segments selected at qualities $q_1(r_i, r)$ and $q_2(r_j, r)$  as:
$$v_i=
\begin{cases}
	\eta (q_{i} - q_{i-1}), & \mbox{if } q_{i} \ge q_{i-1}\\
	-\gamma (q_{i} - q_{i-1}) & \mbox{if } q_{i} < q_{i-1}
\end{cases}$$
where $\eta$ and $\gamma$ positive factors that determines how much changes impact the overall experience when a transition to a higher or to a lower quality representation happens.

We finally formulate our quality model as the objective function capturing the defined values, where for a sequence of N segments selected at qualities $\bm{q}_{1}^N$:
$$\phi(\bm{q}_{1}^N) = \sum_{k=1}^N(q_k-v_k)$$
where $v_k=v(\bm{q}_{k-M}^k)$.

\subsubsection{Cost Factors Considerations}
We can include other metrics in our model to account for the network incentives; for instance, a network operator may have a peering agreement such that it incurs a cost linear with the amount of bandwidth. Therefore we can introduce a cost $c_{bw} = \eta r_i$ with $\eta$ some price per unit of data transfer; also, there may be a cost associated with the storage of data at the edge cache. Namely, if $\omega_s(t)$ is the amount of data in transit from the servers to the clients which is stored on the edge cache at time $t$, we can introduce a storage cost $c_s = \kappa \sum_t \omega_s(t)$. 

We can then subtract these costs from the utility to get a joint objective which takes into account the QoE of the end-user (or the benefit to the end-user as a potential revenue to the operator) minus the costs associated with the system.

\subsection{Rate and Streaming Algorithms}\label{sec:algorithms}

We model our system, introduced in Section \ref{sec:networkModel}, as a composition of four main blocs as shown in Figure \ref{fig:network_model}: a web server, that has complete availability of all the segments composing the video; a long distance network, an access network that uses a caching system in order to support the streaming process; finally, a mobile client attached to the network through a wireless interface. While more than one access interface could be considered, we first focus our analysis on a single access interface. Resources available on the two hops between the server and the client are regulated by the available capacity in the core network for the path connecting the server to the access network where the caches are located, which we will name the core network represented by a function $c(t)$, and by the throughput available to the client for the wireless link giving the client access to the network, which we will name the edge network represented by a function $e(t)$.

Let us first introduce a supporting example to build intuition behind the presented model. In contrast to current solutions where the logic employed to perform the bitrate adaptation process is based on estimation of available resources perceived in the recent past, we consider a partially anticipative case in which a finite window of future edge and core network capacity variations are known beforehand. This information is used by the network controller to schedule which video segments to download in the upcoming future, by either transferring directly the content from the server to the client, or by using the available caches as support.

\begin{figure}[!t]
	\centering
        \begin{subfigure}[b]{0.48\linewidth}
        		\includegraphics[width=1.1\linewidth]{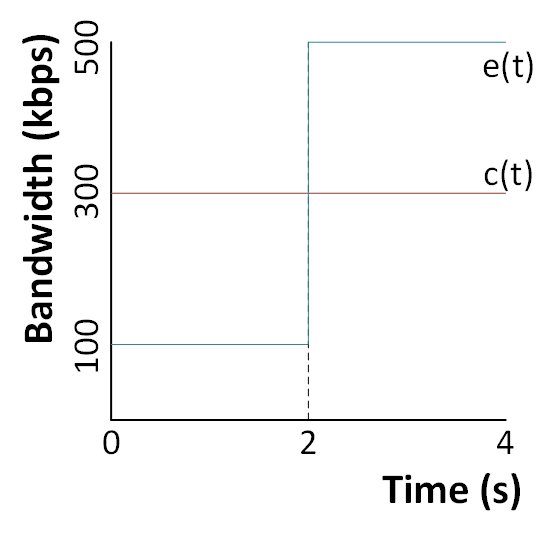}
		\caption{}
                \label{fig:graphA}
        \end{subfigure}~
        	\begin{subfigure}[b]{0.48\linewidth}
        		\includegraphics[width=1.1\linewidth]{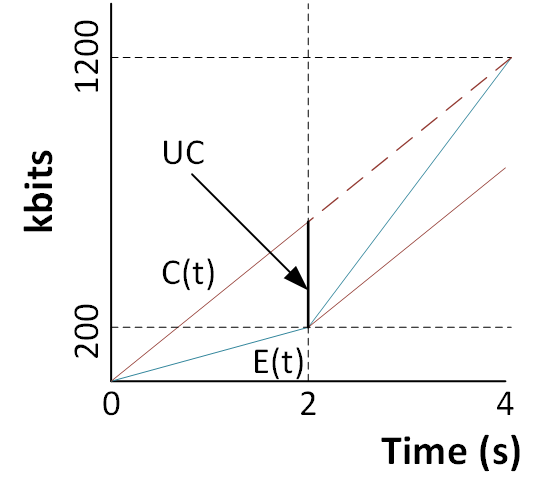}
                \caption{}
                \label{fig:graphB}
        \end{subfigure}
	\caption{Example on how future knowledge can be used to exploit residual capacity. \textbf{(a)} Provides the general scenario and \textbf{(b)} its representation through cumulative downloaded data.}\label{fig:graph}
	
\vspace{-5mm}
\end{figure}

Assume that at time $t$, we know the evolution of the available capacities $c_{[t,t+W]}$ and $e_{[t,t+W]}$, where $W$ is the future knowledge window size, and we do not know the capacity beyond $t+W$. Consider the simple scenario represented in Figure \ref{fig:graphA}, where $W=4$ and $c(t)$ is constant at 300kbps and $e(t)$ varies from a first 2 seconds period at 100kbps to a second period at 500kbps. In this context a client wants to retrieve a video divided in segments of size 2 seconds and available at three different representations: 100kbps, 300kbps and 600kbps. Initially the video buffer is empty, so let us assume that the first segment of video will be downloaded at minimum quality to reduce the startup wait time for the client. This corresponds to downloading 200kbits of data which in our example, given the initial bottleneck of the wireless link will take 2 seconds to complete. Once this happens, the video client has 2 seconds of video available for display and it should try to download the next segment by this amount of time otherwise a rebuffering event would occur (i.e. the video client remains stuck waiting for more video data to be available). We consider the amount of time until the moment the next downloaded segment will be displayed its download deadline $t_{d}$. $t_{d}$ in our example will then occur at the time of 4 seconds. In current scenarios, where downloads are controlled by the client, only two representations would meet the deadline: 100kbps (which would allow for more segments to be downloaded) and 300kbps. In our anticipative case, we know that even though the first download will be bandwidth limited by the current bottleneck, we are actually under-utilizing the network resources as 200kbps of unused capacity have not been exploited during the first segment download. As we are scheduling downloads in advance, when evaluating the amount of time needed to download next segment we consider the unused capacity as data that might actually have been downloaded to the local caches. The actual amount of data that could have been downloaded to a location in the network (i.e. the caches) is easily calculated from considering the cumulative versions of $e(t)$ and $c(t)$, that we will call $E(t)$ and $C(t)$, as:
$$UC(t) = C(t) - E(t)$$
In our particular case, $UC(t)$ is highlighted in black in Figure \ref{fig:graphB} and corresponds to 400kbits. Now we can assume that this amount of data could have already been transferred close to the edge network preventing the core network from being the bottleneck in the second period. This difference can again be noticed in Figure \ref{fig:graphB}: given that the amount of downloaded data always corresponds to the minimum of the two functions $E(t)$ and $C(t)$ (as a client can never download more than what is allowed by the bottleneck in the network), at the time of 2 seconds this would correspond to 200kbits. Assuming no preemptive caching is applied, the amount of data that could be downloaded in the second period is again delimited by the minimum of the two lines and in this case it would correspond to the plain line representing $C(t)$. If we assume that the core network has moved to the caches an amount of data corresponding to the unused capacity, the minimum has now to be taken between $E(t)$ and the dotted version of $C(t)$. 

Our algorithm applies these concepts of deadlines and evaluations of unused capacity to explore the state space of valid combinations of segment bitrate downloads to select the one in the time window that produces the highest QoE utility value described in Section \ref{sec:qoeModel} without ever occurring in rebuffering events (hence always meeting the set deadlines). 

\vspace{2mm}\noindent\textbf{Bitrate Selection Algorithm} We now formalize the provided concepts; we define the streaming process as a combination of two tasks for each window of future knowledge: first the controller \textbf{schedules segments downloads} given the available predicted resources and the client reproduction status; second the  \textbf{delivery of the segments} to the the mobile hosts with use of the edge caches.
The core of the scheduling algorithm is based on a recursive function performed at the beginning of each time window as described in Algorithm \ref{alg:spacec}; this function searches for an optimal path (sequence of segment bitrates) among all possible combinations. Starting from time $t$ and given a potential bitrate $j$ for the next required segment with index $i$ determines the effect of such representation on the download process and, once consumed the entire future knowledge window, determines the QoE of the selected path. The function calculates the download time for the given segment using the available throughput and residual capacity from previous steps. Each time the function is called from the base algorithm, the starting residual capacity is assumed to be zero. In recursive calls, the residual capacity in the core network, if available according to the known cumulative throughput functions $C$ and $E$, may be adjusted. As long as the knowledge window limit is not reached, the recursion follows. Once reached the end of the window (or potentially the end of the video), the utility function value of the current path is calculated and compared with the best path previously found and only the better of the two is kept. In case a rebuffering time event is detected, the path is declared invalid and the function returns. Our algorithm can be summarized in four main steps:
\begin{enumerate}
\item Avoid overrunning the client buffer by eventually waiting until some space is created (lines 9 to 17).
\item Calculate the download deadline for the considered segment ($t_{i,d}$) and evaluate given the previously accumulated unused capacity if the deadline can be met by calculating the amount of time necessary to transfer the required data ($t_{i,e}$); if not return the previously found best path (lines 18 to 24).
\item If the future window has been consumed or the end of the file is reached return the $GREATEST$ between the current path and the previously best path based on their utility value (lines 25 to 27 and lines from 35 to 37).
\item Otherwise recursively evaluate the same function for the next segment (index $i+1$) over its possible representations. (lines 28 to 34).
\end{enumerate}
After completing the recursive process, the sequence of segments with the highest QoE value is returned and the controller can use it to instruct the other components on how to proceed (i.e. instruct the clients and the caches on how to proceed for downloading the selected segments).

\begin{algorithm}[t]
\begin{algorithmic}[1]
\Function{findOptimalPath}{$i, j, UC, W_r, B, D_t, \phi_{best}$}
\State \emph{// $i$ - next segment index}
\State \emph{// $j$ - assumed bitrate of the next segment}
\State \emph{// $UC$ - available residual capacity}
\State \emph{// $W_r$ - remaining time in knowledge window}
\State \emph{// $B$ video buffer available at client}
\State \emph{// $D_t$ time to complete displayed segment}
\State \emph{// $\phi_{best}$ - current best path found}
\If {\emph{B} $==$ \emph{max\_buffer\_size}}
\State wait until the end of current displayed segment
\If {$D_t > W_r$}
\State \Return \Call{greatest}{$\phi_{best}$, \emph{current\_path}}
\Else
\State \text{reduce $B$ by segment size}
\State $D_t \gets \text{segment size}$
\EndIf
\EndIf
\State $UC \gets$ increment given waited time
\State $t_{i,e}\gets$ segment download time
\If {$t_{i,e} < Wr$}
\If {$t_{i,e} > B + D_t$}
\State \emph{// Rebuffering time \textgreater\ 0}
\State \Return $\phi_{best}$
\EndIf
\If {$i == N$}
\State \emph{// Reached end of video}
\State \Return \Call{greatest}{$\phi_{best}$, \emph{current\_path}}
\Else
\State $W_r \gets W_r - t_{i,e}$
\State update $B$, $D_t$ and $UC$ given $t_{i,e}$
\ForAll{bitrates $m$}
\State \Call{findOptimalPath}{${i+1},m,UC,W_r,B,D_t,\phi_{best}$}
\EndFor
\EndIf
\Else
\State \Return \Call{greatest}{$\phi_{best}$, \emph{current\_path}}
\EndIf 
\EndFunction
\end{algorithmic}
\caption{Path building algorithm}
\label{alg:spacec}
\end{algorithm}

\vspace{2mm}\noindent\textbf{Interleaving windows.} The main characteristic of our algorithm is that it greedily tries to use as many resources as are available in the time window without much consideration for the following time slots. Since the QoE cost of switching to a higher bitrate is lower than the cost of switching to lower bitrates, the consequence of such behavior is the tendency to select higher bitrates toward the end of the window to consume the remaining capacity available. This is not always the optimal path in the longer run, since it might be necessary to choose a lower bitrate at the beginning of the next time window, and thus suffer the QoE drop due to switching to a lower bitrate immediately afterwards. To avoid this problem, we consider an alternative solution: while we still apply the same algorithm for the complete window of size $W$ to select the best path, we only apply the obtained optimal path until an earlier moment $W - t_i$, where $t_i$ is smaller than $W$. This way we prevent a higher bitrate from being selected in the last $t_i$ of the time window $W$, avoiding possible quality drops at the beginning of the next window. After this is done, the new considered window will start from time $W - t_i$. 

\vspace{2mm}\noindent\textbf{Cost analysis.} The cost of the algorithm can be exponential if the bandwidth considered is always bigger than video bitrate (i.e. all possible representations could be downloaded). While this obviously is an obstacle towards the deployment of this algorithm, a set of simple measures and considerations can be taken into account to have an effectively lesser actual cost in real deployments: (1) as quality transitions (in particular negative ones) negatively affect the final utility value, we can set a limit on the number of these events; (2) if the bottleneck bandwidth is always higher than a certain bitrate during the duration of a time window, all video representations with lower bitrate can be left out of consideration (3) in most cases the total number of downloaded segments will be limited as the window size is limited.

\section{Simulations}\label{sec:experiments}
In this section, we evaluate the gains achieved by our system and joint rate adaptation through a set of MATLAB based simulations. In order to understand its potential, we implemented the core logic of our system and compared it to the behavior of common DASH implementations. While different proprietary algorithms are used in some of the available commercial solutions (e.g. Apple HLS, Adobe HDS, etc.), we implement our baseline following the behavior in the logic implemented by Netflix-like video services, which can be summarized in two main characteristics: 
\begin{itemize}
\item the DASH client downloads and keeps only video segments for the following $t$ seconds of playback at any given time (i.e. the buffer size is limited by time, not data space);
\item the DASH client logic adapts video quality by a moving average of the data rate estimates experienced on the previous $k$ segments delivered (we set $k$ to 5 for our experiments).
\end{itemize}
While it is easy to identify a wide selection of factors that might affect the final results of our simulations, we try to fairly compare the baseline with the two variants of our algorithm by applying for each run the same conditions (i.e. evolution of network infrastructure resources during the experiment time). In the next two subsections we will describe the model used to characterize these resources and the video data set employed in our tests.

\vspace{2mm}\noindent\textbf{Video Dataset and QoE Model.} For this set of simulations, we used a video content of 5 minutes of length. The video is divided into 2 second long segments, with each segment available in three different bitrate representations (1 Mbps, 400 kbps and 100 kbps for the first case, 2 Mbps, 1.2 Mbps and 300 kbps for the second). While our system supports Variable Bitrate for the the video segments, we use only constant bitrates for these simulations (i.e. all segments at the same quality level have the same size). The Quality of Experience is calculated following the description provided in Section \ref{sec:qoeModel}, using parameters: $\alpha = 1$, $\beta = 1$, $\gamma = 1$ = $\eta = 0.1$.

\vspace{2mm}\noindent\textbf{System Resources and Network Model.} In our simulations we do not take into consideration availability of video segments at different servers; we use a single server that has the desired video content available at all times. Moreover, we do not consider any limit in the cache size of the intermediate nodes. We provide results using two different network models:
\begin{itemize}
\item first, we model the available network bandwidth as two finite-state, discrete-time Markov chains, where transitions occur at constant times, every 2 seconds, and transitions occur only between the two nearest states; this is done to try to capture slow variations attributable to client mobility (for the wireless links) and evolution in congestion for the core network. These Markov chains are:
\begingroup\makeatletter\def\f@size{8}\check@mathfonts
\def\maketag@@@#1{\hbox{\m@th\large\normalfont#1}}%
$$
P_W = \begin{bmatrix} 
0.5& 0.5& 0.0& 0.0& 0.0& 0.0\\
0.33& 0.33& 0.33& 0.0& 0.0& 0.0\\
0.0& 0.33& 0.33& 0.33& 0.0& 0.0\\
0.0& 0.0& 0.33& 0.33& 0.33& 0.0\\
0.0& 0.0& 0.0& 0.4& 0.4& 0.2\\
0.0& 0.0& 0.0& 0.0& 0.7& 0.3
\end{bmatrix}
R_W = \begin{bmatrix}
100\\ 300\\ 500\\ 700\\ 900\\ 2300
\end{bmatrix}
$$
$$
P_C = \begin{bmatrix} 
0.5& 0.5& 0.0& 0.0& 0.0& 0.0\\
0.33& 0.33& 0.33& 0.0& 0.0& 0.0\\
0.0& 0.33& 0.33& 0.33& 0.0& 0.0\\
0.0& 0.0& 0.33& 0.33& 0.33& 0.0\\
0.0& 0.0& 0.0& 0.33& 0.33& 0.33\\
0.0& 0.0& 0.0& 0.0& 0.5& 0.5
\end{bmatrix}
R_C = \begin{bmatrix}
700\\ 900\\ 1100\\ 1100\\ 1100\\ 1300
\end{bmatrix}
$$
\endgroup
where  $P_W$ represents the transition matrix for the wireless access network where for each state the corresponding bandwidth is shown in $R_W$ (expressed in kbps); the same values are shown for the core network in $P_C$ and $R_C$.
\item second, we use a real cellular trace, collected and presented in \cite{riiser2013commute}. This trace consists of multiple days worth of data collected in a real metropolitan environment in a European city, using a HSDPA modem while commuting with public transportation. The core network is modeled similarly to the previous case, using the same transition matrix but with the following values for the available throughput: 
$R_C = [1000; 1200; 1400; 1600; 1800; 2000]^T$.
\end{itemize}

\begin{figure*}
        \centering
        \begin{subfigure}[b]{0.25\textwidth}
        		\includegraphics[width=\linewidth]{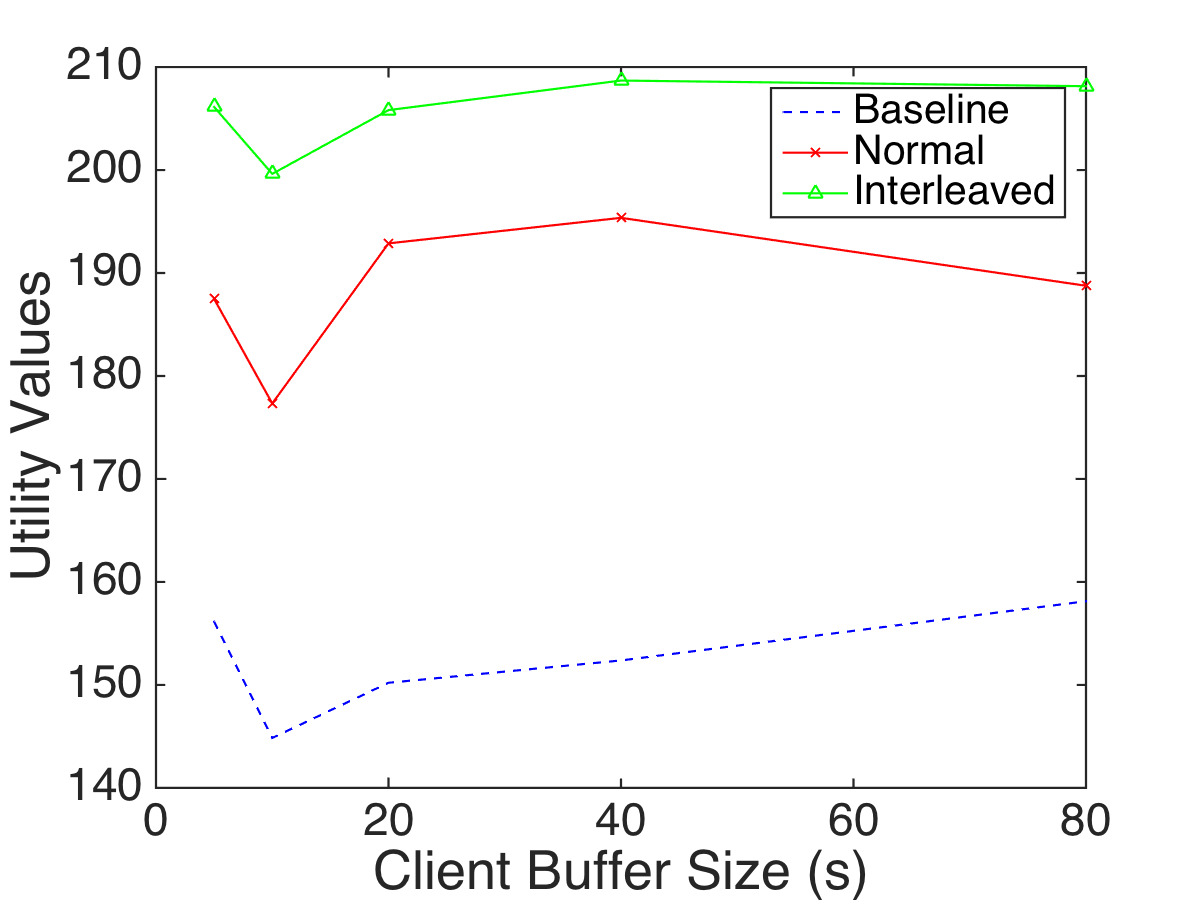}
                \caption{Utility value with varying buffer size}
                \label{fig:resa}
        \end{subfigure}%
        ~ 
        \begin{subfigure}[b]{0.25\textwidth}
        		\includegraphics[width=\linewidth]{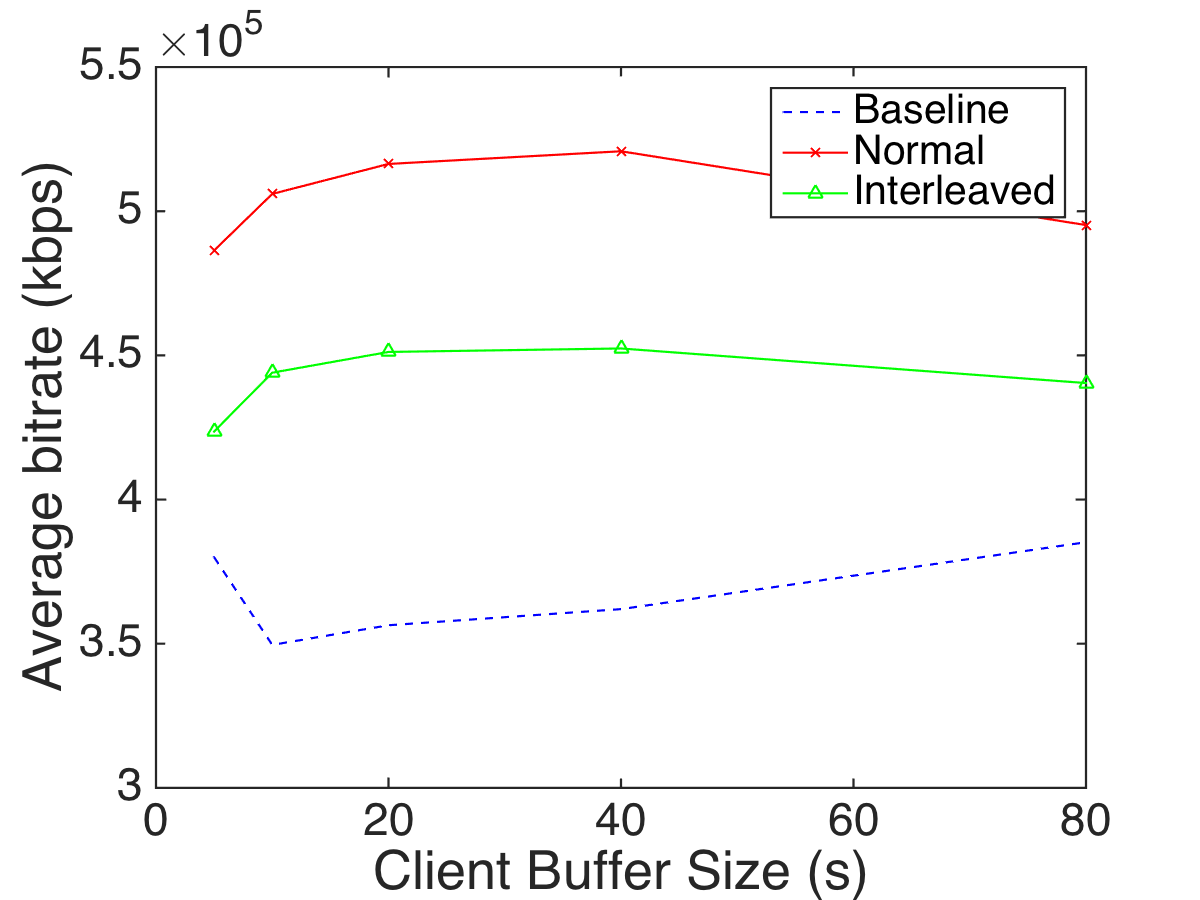}
                \caption{Average bitrate with varying buffer size}
                \label{fig:resb}
        \end{subfigure}%
        ~ 
        \begin{subfigure}[b]{0.25\textwidth}
        		\includegraphics[width=\linewidth]{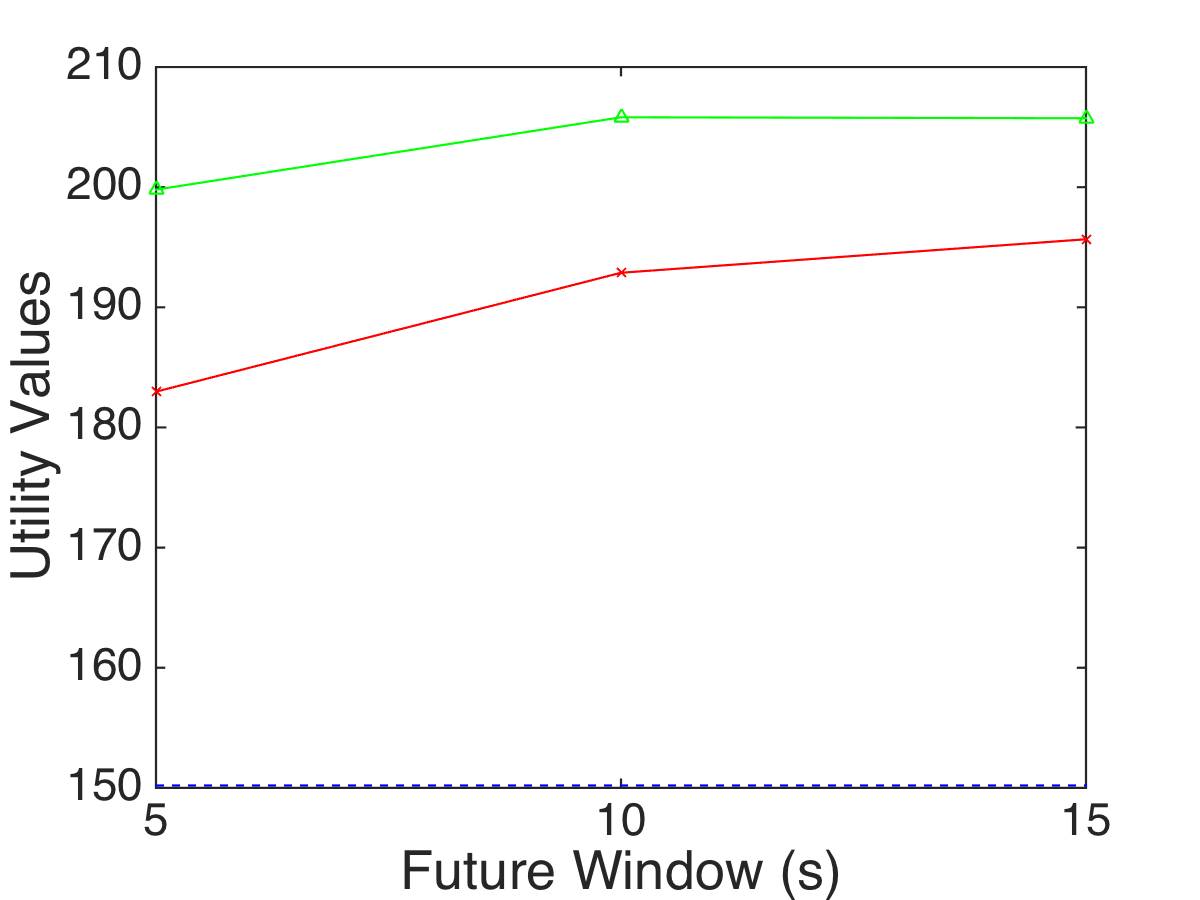}
                \caption{Utility value with varying future window size}
                \label{fig:resc}
        \end{subfigure}%
        ~ 
        \begin{subfigure}[b]{0.25\textwidth}
        		\includegraphics[width=\linewidth]{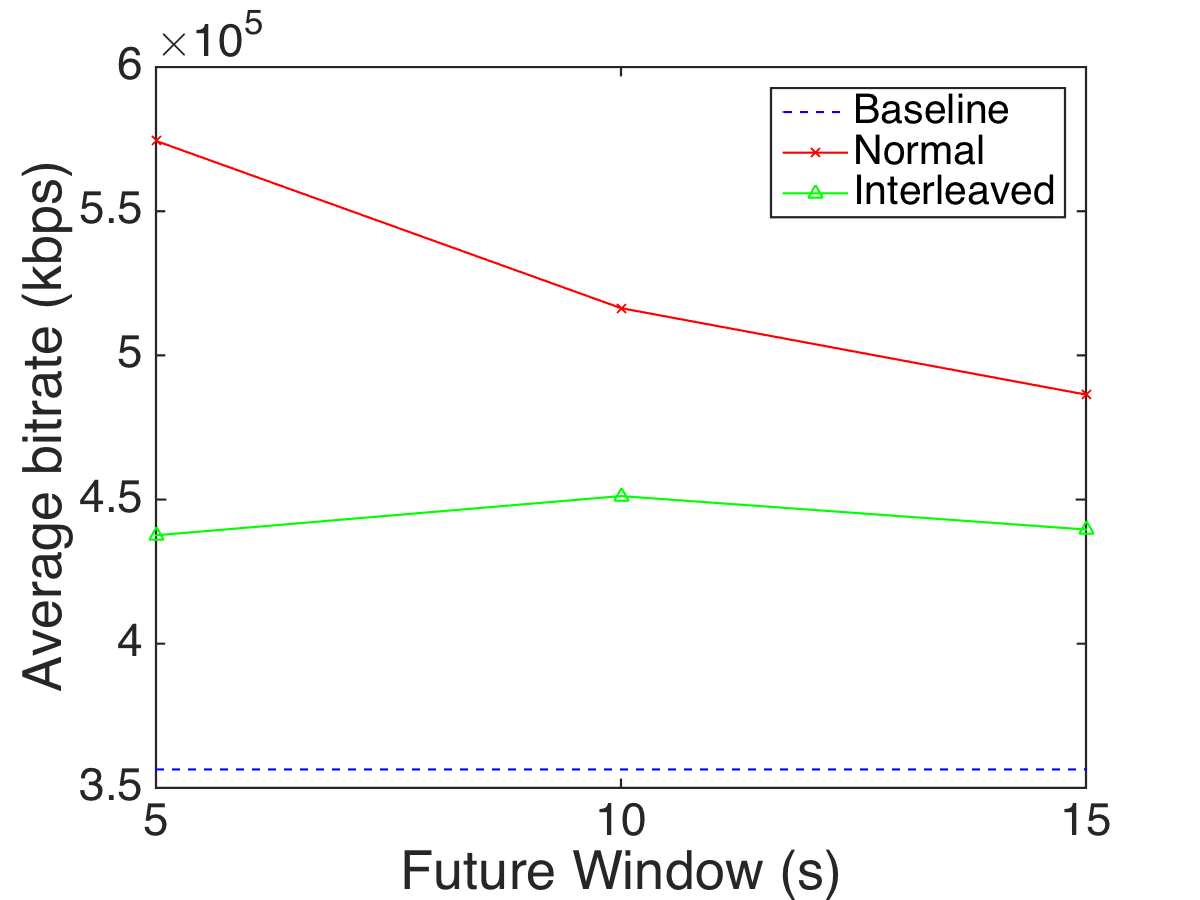}
                \caption{Average bitrate with varying future window size}
                \label{fig:resd}
        \end{subfigure}
        \\
        \centering
        \begin{subfigure}[b]{0.25\textwidth}
        		\includegraphics[width=\linewidth]{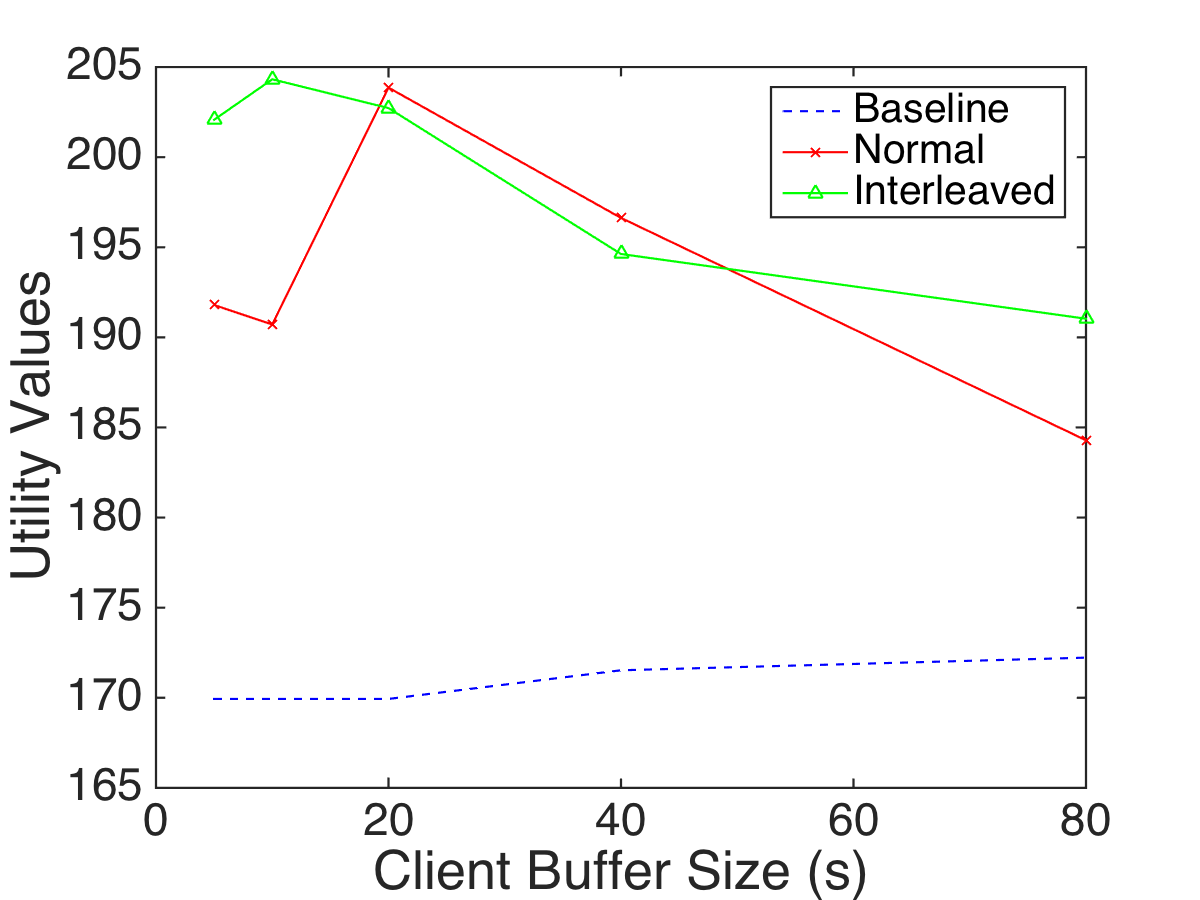}
                \caption{Utility value with varying buffer size}
                \label{fig:resa}
        \end{subfigure}%
        ~ 
        \begin{subfigure}[b]{0.25\textwidth}
        		\includegraphics[width=\linewidth]{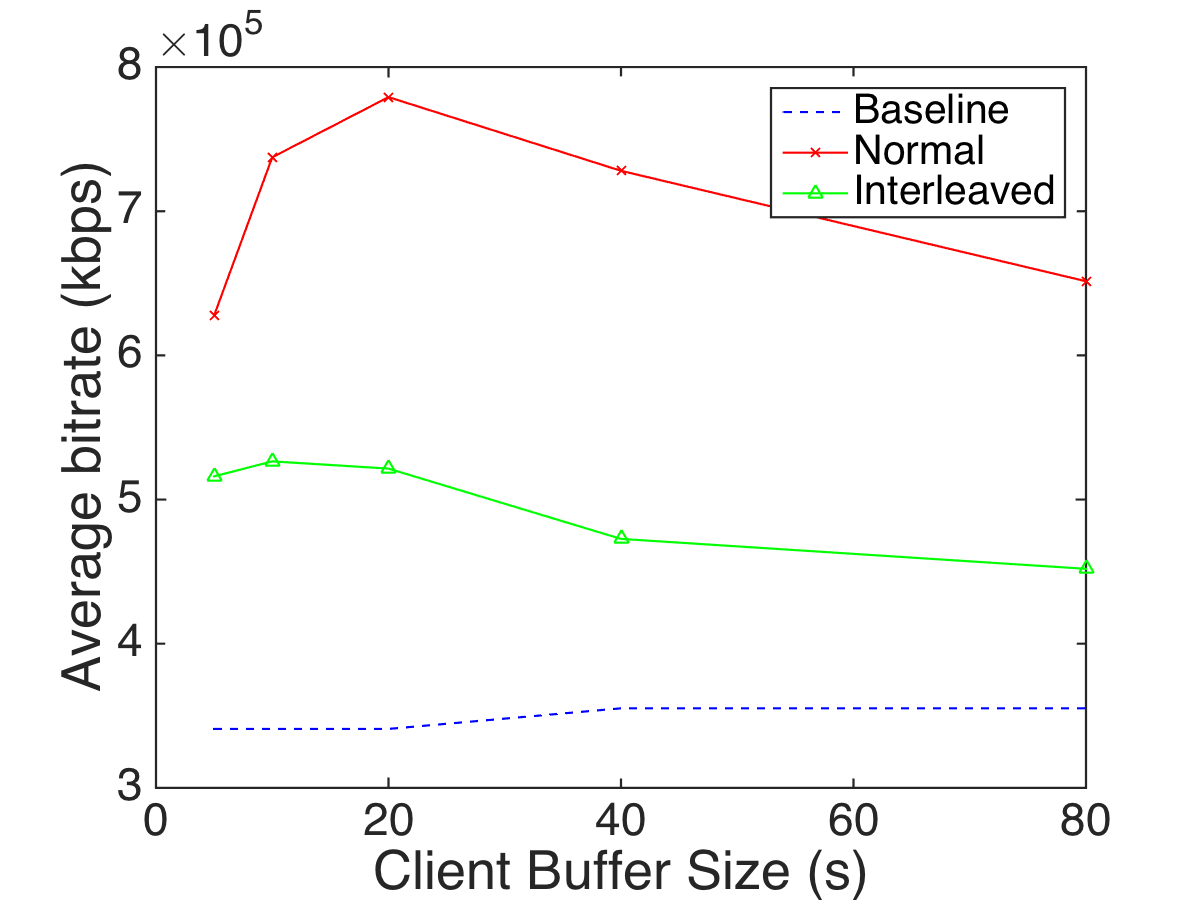}
                \caption{Average bitrate with varying buffer size}
                \label{fig:resb}
        \end{subfigure}%
        ~ 
        \begin{subfigure}[b]{0.25\textwidth}
        		\includegraphics[width=\linewidth]{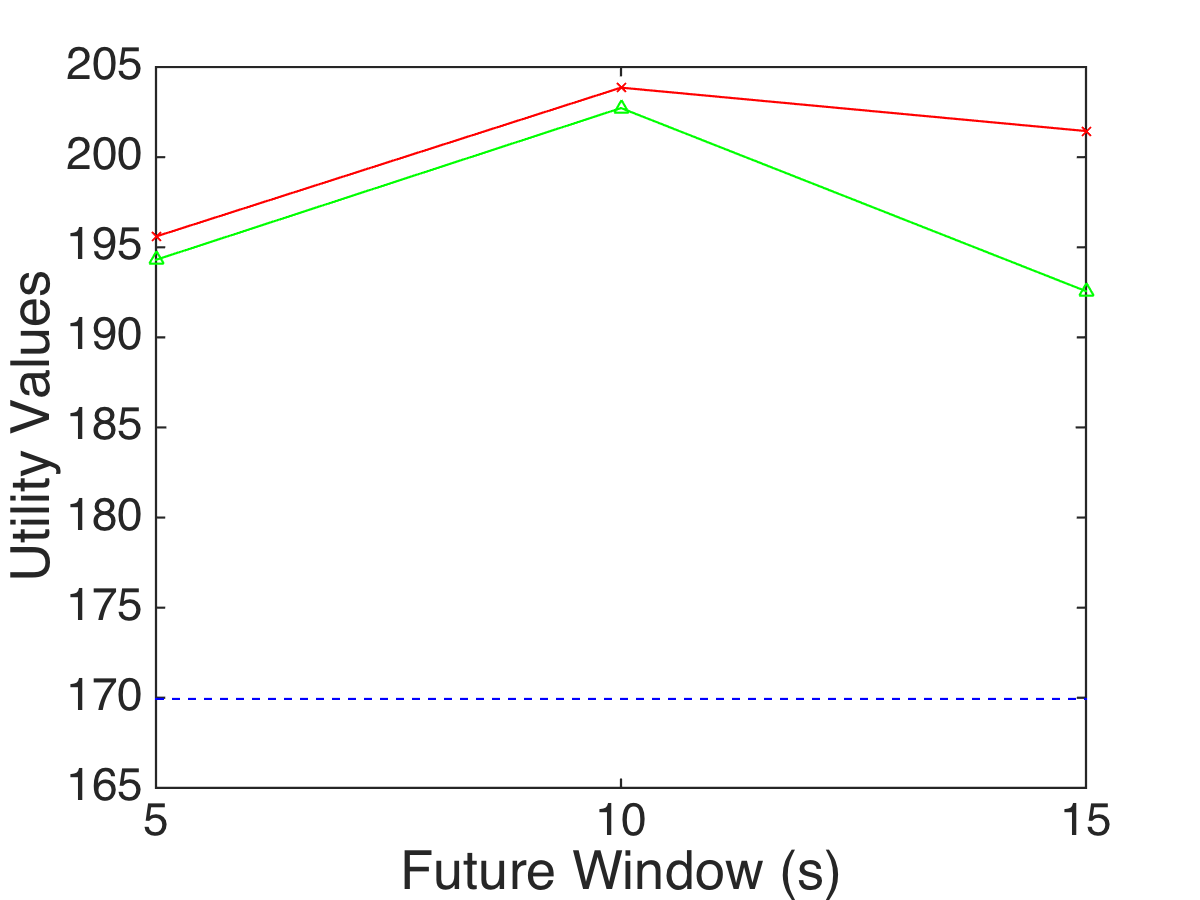}
                \caption{Utility value with varying future window size}
                \label{fig:resc}
        \end{subfigure}%
        ~ 
        \begin{subfigure}[b]{0.25\textwidth}
        		\includegraphics[width=\linewidth]{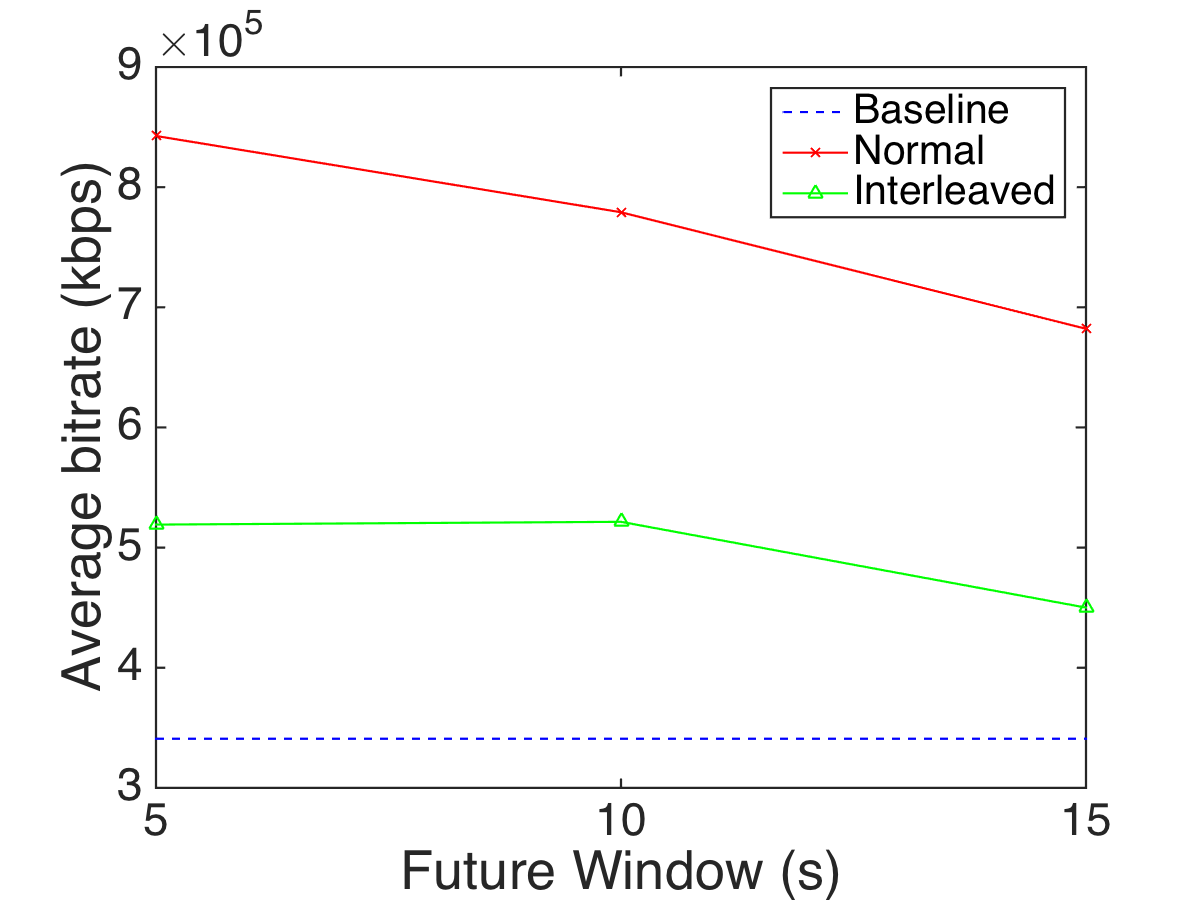}
                \caption{Average bitrate with varying future window size}
                \label{fig:resd}
        \end{subfigure}
        \caption{Simulation results. Figures a-d provide results for a Markov chain based wireless model, while Figures e-h use real world traces collected in an urban environment.}\label{fig:results}
        
\vspace{-5mm}
\end{figure*}

\vspace{2mm}\noindent\textbf{Results.} We evaluate our system under two varying factors: video buffer size available to the DASH client and window of future knowledge of available bandwidth at the two analyzed network components. Figure \ref{fig:results} collects all the results obtained. For each of the data points represented, we repeated the experiment 5 times and collected the average result. This does not apply for the baseline in Figures \ref{fig:resc} and \ref{fig:resd} as the future window size can vary only for our algorithms; for these cases we used a single data point using client buffer size of 20 seconds. In general, the results confirm the overall benefit of our system with gains of at least 15\% points in the utility value for both our algorithm versions. This not only corresponds to a more stable experience (as variations, in particular decreases in quality, strongly affect the final QoE value), but also in a higher average bitrate quality for all the experiments analyzed.

As expected, the buffer size available for the clients is not a major influencing factor for the analyzed use case, as neither algorithm uses the buffer size information to modify its adaptation logic; we expect this parameter to gain more importance for longer videos, where a low bandwidth period might be better compensated by accumulated buffer.

More interestingly, we can notice that increasing future knowledge window size also increases the computational cost of schedule delivery time deadlines, since there are now longer periods of time that are considered in the calculation. The achieved gains in this case might not justify the additional processing time costs.

\begin{figure}[h!]
        \centering
        \begin{subfigure}[b]{0.45\linewidth}
        		\includegraphics[width=\linewidth]{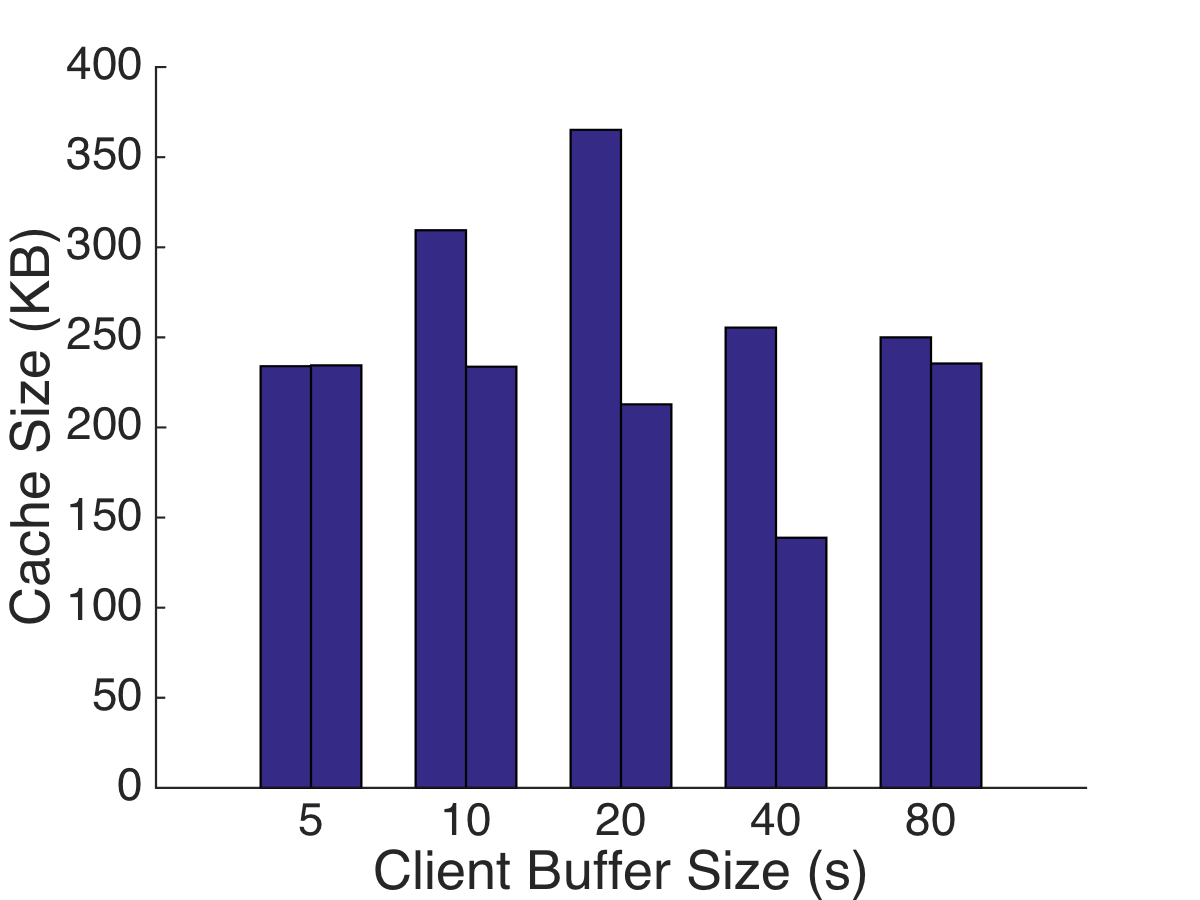}
                \caption{}
                \label{fig:cache_alg1}
        \end{subfigure} ~
        \begin{subfigure}[b]{0.45\linewidth}
        		\includegraphics[width=\linewidth]{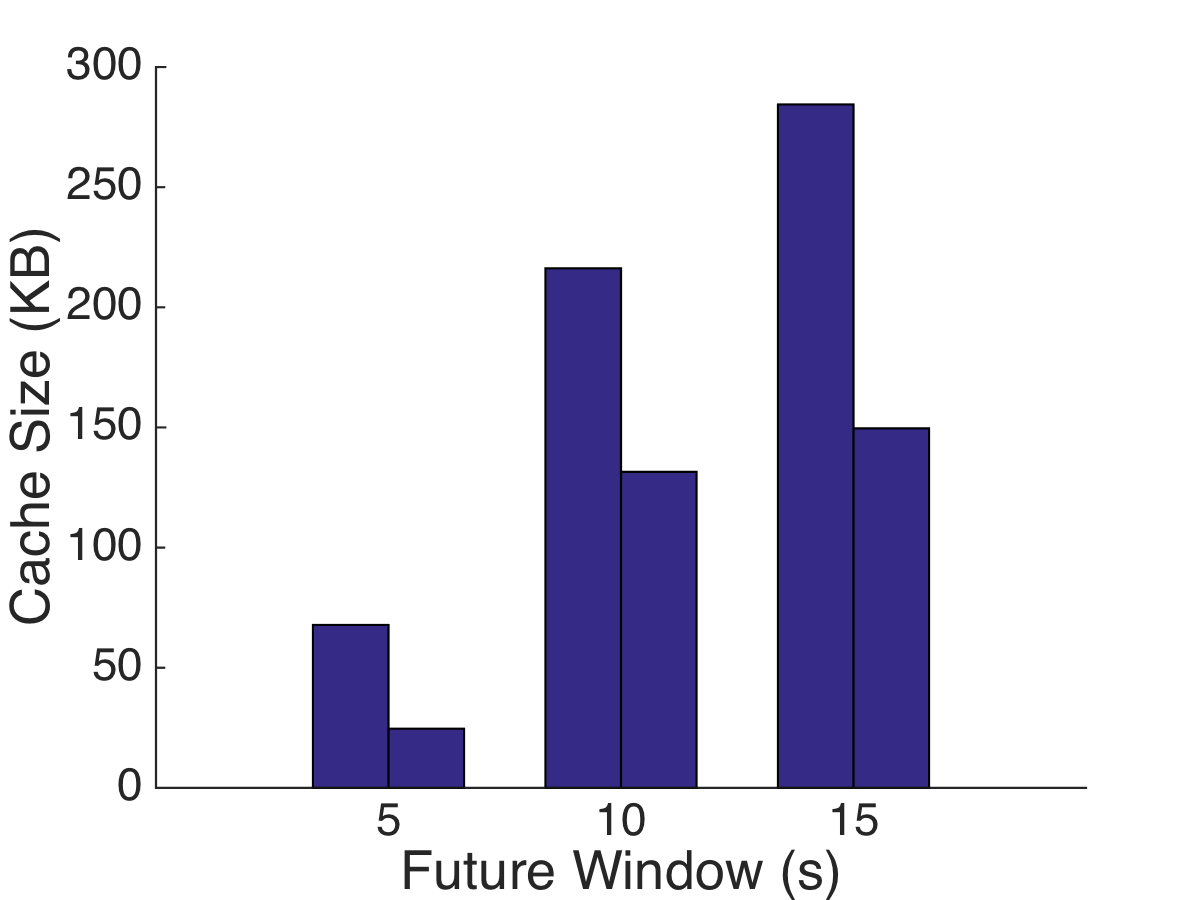}
                \caption{}
                \label{fig:cache_alg2}
        \end{subfigure}
        \caption{Max buffer occupancy over different runs.}\label{fig:cache_results}
\vspace{-5mm}
\end{figure}

Figure \ref{fig:cache_results} shows the employed memory usage at the network caches. In particular, it shows the cache size for the single client experiment of the first analyzed case. This value enables us to understand the resource cost of running our system, and is of particular interest to service providers that would implement such a caching scheme as scalability is directly affected by the consumed resources. The amount of video data buffered at the cache nodes using our algorithm ranged from $0$ up to $365 kB$, meaning that it would scale well for larger videos and higher number of clients and not impose a high resource cost for the service provider.



\section{Conclusions}\label{sec:conclusions}

In order to optimize delivering video content to clients, we have incorporated the knowledge of network throughput in a short future time window. Our model is based on DASH, that is now used in most major video streaming services, and in comparison with DASH it improves the usage of the available end-to-end throughput between the server and the client. Instead of having the entire network throughput equal to the bottleneck of the used network path, we optimize the video delivery method, utilizing the intermediate nodes between the server that has the whole video content available, and the client, as an intermediate video content cache. Using this cache enables us to increase the effective throughput in certain parts of the network path. In times when the core network has higher throughput than the edge network, we increase the cache, and then use the cached content in periods when the edge network is faster.

We evaluated the gains achieved by using our method with a set of MATLAB simulations. We implemented our algorithm and compared it with the behavior of DASH implementations against synthetic and real throughput traces.  Our evaluations demonstrate that our method of video delivery improves the Quality of Experience for the end user by keeping the bitrate more stable and achieving significantly higher average bitrates than the benchmarks. While it does introduce some additional costs for the content provider, we claim that the resource costs are minimal, and the achieved gain outweighs the computational costs.




\bibliographystyle{unsrt}
\bibliography{references}

\begin{thebibliography}{10}

\bibitem{ciscoForecast}
{Cisco Visual Networking Index: Forecast and Methodology, 2013–2018}.
\newblock
  \url{http://www.cisco.com/c/en/us/solutions/collateral/service-provider/ip-ngn-ip-next-generation-network/white_paper_c11-481360.html/}.

\bibitem{bianchi1997role}
Giuseppe Bianchi and Riccardo Melen.
\newblock The role of local storage in supporting video retrieval services on
  atm networks.
\newblock {\em IEEE/ACM ToN}, 5(6), 1997.

\bibitem{sodagar2011mpeg}
Iraj Sodagar.
\newblock The mpeg-dash standard for multimedia streaming over the internet.
\newblock {\em MultiMedia, IEEE}, 18(4):62--67, 2011.

\bibitem{jacobson2007content}
Van Jacobson, Marc Mosko, D~Smetters, and JJ~Garcia-Luna-Aceves.
\newblock Content-centric networking.
\newblock {\em Whitepaper, Palo Alto Research Center}, pages 2--4, 2007.

\bibitem{dannewitz2009netinf}
Christian Dannewitz.
\newblock Netinf: An information-centric design for the future internet.
\newblock In {\em Proc. 3rd GI/ITG KuVS Workshop on The Future Internet}, 2009.

\bibitem{koponen2007data}
Teemu Koponen, Mohit Chawla, Byung-Gon Chun, Andrey Ermolinskiy, Kye~Hyun Kim,
  Scott Shenker, and Ion Stoica.
\newblock A data-oriented (and beyond) network architecture.
\newblock In {\em ACM SIGCOMM Computer Communication Review}, volume~37, pages
  181--192. ACM, 2007.

\bibitem{ledereradaptive}
C~Westphal~(ed).
\newblock Adaptive streaming over icn, irtf icnrg wg,
  draft-irtf-icnrg-videostreaming-02.

\bibitem{grandl2013interaction}
Reinhard Grandl, Kai Su, and Cedric Westphal.
\newblock On the interaction of adaptive video streaming with content-centric
  networking.
\newblock {\em arXiv preprint arXiv:1307.0794}, 2013.

\bibitem{lee2013svc}
Junghwan Lee, Jaehyun Hwang, Nakjung Choi, and Chuck Yoo.
\newblock Svc-based adaptive video streaming over content-centric networking.
\newblock {\em KSII Transactions on Internet and Information Systems (TIIS)},
  7(10):2430--2447, 2013.

\bibitem{malandrino2012proactive}
Francesco Malandrino, Maciej Kurant, Athina Markopoulou, Cedric Westphal, and
  Ulas~C Kozat.
\newblock Proactive seeding for information cascades in cellular networks.
\newblock In {\em IEEE INFOCOM'12}, pages 1719--1727. IEEE, 2012.

\bibitem{ariyasinghe2013distributed}
LR~Ariyasinghe, C~Wickramasinghe, PMAB Samarakoon, UBP Perera, RA~Prabhath
  Buddhika, and MN~Wijesundara.
\newblock Distributed local area content delivery approach with heuristic based
  web prefetching.
\newblock In {\em IEEE ICCSE'13}, pages 377--382, 2013.

\bibitem{lu2013optimizing}
Zheng Lu and Gustavo de~Veciana.
\newblock Optimizing stored video delivery for mobile networks: the value of
  knowing the future.
\newblock In {\em INFOCOM, 2013 Proceedings IEEE}, pages 2706--2714. IEEE,
  2013.

\bibitem{balachandran2013analyzing}
Athula Balachandran, Vyas Sekar, Aditya Akella, and Srinivasan Seshan.
\newblock Analyzing the potential benefits of {CDN} augmentation strategies for
  internet video workloads.
\newblock In {\em ACM IMC'13}, pages 43--56, 2013.

\bibitem{plissonneau2012longitudinal}
Louis Plissonneau and Ernst Biersack.
\newblock A longitudinal view of http video streaming performance.
\newblock In {\em ACM MSC'13}, pages 203--214. ACM, 2012.

\bibitem{finamore2011youtube}
Alessandro Finamore, Marco Mellia, Maurizio~M Munaf{\`o}, Ruben Torres, and
  Sanjay~G Rao.
\newblock Youtube everywhere: Impact of device and infrastructure synergies on
  user experience.
\newblock In {\em ACM IMC'11}, pages 345--360. ACM, 2011.

\bibitem{li2012network}
Zhe Li, Mohamed~Karim Sbai, Yassine Hadjadj-Aoul, Annie Gravey, Damien Alliez,
  Jeremie Garnier, Gerard Madec, Gwendal Simon, and Kamal Singh.
\newblock Network friendly video distribution.
\newblock In {\em International Conference on the Network of the Future},
  volume~1, 2012.

\bibitem{liu2013peer}
Fangming Liu, Bo~Li, and H~Jin.
\newblock Peer-assisted on-demand streaming: characterizing demands and
  optimizing supplies.
\newblock {\em IEEE Trans on Computers}, 2013.

\bibitem{liu2012optimizing}
Hongqiang~Harry Liu, Ye~Wang, Yang~Richard Yang, Hao Wang, and Chen Tian.
\newblock Optimizing cost and performance for content multihoming.
\newblock {\em ACM SIGCOMM CCR}, 42(4):371--382, 2012.

\bibitem{liu2012case}
Xi~Liu, Florin Dobrian, Henry Milner, Junchen Jiang, Vyas Sekar, Ion Stoica,
  and Hui Zhang.
\newblock A case for a coordinated internet video control plane.
\newblock In {\em ACM SIGCOMM'12}, pages 359--370, 2012.

\bibitem{huang2008understanding}
Cheng Huang, Angela Wang, Jin Li, and Keith~W Ross.
\newblock Understanding hybrid cdn-p2p: why limelight needs its own red swoosh.
\newblock In {\em ACM NOSSDAV'08}, pages 75--80. ACM, 2008.

\bibitem{huang2007can}
Cheng Huang, Jin Li, and Keith~W Ross.
\newblock Can internet video-on-demand be profitable?
\newblock In {\em ACM SIGCOMM CCR}, volume~37, pages 133--144. ACM, 2007.

\bibitem{jin2003network}
Shudong Jin, Azer Bestavros, and Arun Iyengar.
\newblock Network-aware partial caching for internet streaming media.
\newblock {\em Multimedia systems}, 9(4):386--396, 2003.

\bibitem{miao1999proxy}
Zhourong Miao and Antonio Ortega.
\newblock Proxy caching for efficient video services over the internet.
\newblock In {\em Packet Video Workshop (PVW'99)}, 1999.

\bibitem{li2005cache}
Keqiu Li, Keishi Tajima, and Hong Shen.
\newblock Cache replacement for transcoding proxy caching.
\newblock In {\em IEEE/WIC/ACM Web Intelligence'05}, pages 500--507, 2005.

\bibitem{qu2005cache}
Wenyu Qu, Keqiu Li, Hong Shen, Yingwei Jin, and Takashi Nanya.
\newblock The cache replacement problem for multimedia object caching.
\newblock In {\em Semantics, Knowledge and Grid, 2005. SKG'05. First
  International Conference on}, pages 26--26. IEEE, 2005.

\bibitem{sen1999proxy}
Subhabrata Sen, Jennifer Rexford, and Don Towsley.
\newblock Proxy prefix caching for multimedia streams.
\newblock In {\em IEEE INFOCOM'99}, volume~3, pages 1310--1319, 1999.

\bibitem{park2001popularity}
Seong-Ho Park, Eun-Ji Lim, and Ki-Dong Chung.
\newblock Popularity-based partial caching for vod systems using a proxy
  server.
\newblock In {\em Proceedings of the 15th International Parallel \& Distributed
  Processing Symposium}, page 115. IEEE Computer Society, 2001.

\bibitem{chen2005segment}
Songqing Chen, Haining Wang, Xiaodong Zhang, Bo~Shen, and Susie Wee.
\newblock Segment-based proxy caching for internet streaming media delivery.
\newblock {\em MultiMedia, IEEE}, 12(3):59--67, 2005.

\bibitem{yan2014qoe}
Zhisheng Yan, Jingteng Xue, and Chang~Wen Chen.
\newblock {QoE continuum driven HTTP} adaptive streaming over multi-client
  wireless networks.
\newblock In {\em Multimedia and Expo (ICME), 2014 IEEE International
  Conference on}.

\bibitem{pu2012video}
Wei Pu, Zixuan Zou, and Chang~Wen Chen.
\newblock Video adaptation proxy for wireless dynamic adaptive streaming over
  http.
\newblock In {\em Packet Video Workshop (PV), 2012 19th International}, pages
  65--70. IEEE, 2012.

\bibitem{krishnamoorthi2013helping}
Vengatanathan Krishnamoorthi, Niklas Carlsson, Derek Eager, Anirban Mahanti,
  and Nahid Shahmehri.
\newblock Helping hand or hidden hurdle: Proxy-assisted http-based adaptive
  streaming performance.
\newblock In {\em Proc. IEEE MASCOTS}, 2013.

\bibitem{stockhammer2011dynamic}
Thomas Stockhammer.
\newblock Dynamic adaptive streaming over http--: standards and design
  principles.
\newblock In {\em Proceedings of the second annual ACM conference on Multimedia
  systems}, pages 133--144. ACM, 2011.

\bibitem{huang2014buffer}
Te-Yuan Huang, Ramesh Johari, Nick McKeown, Matthew Trunnell, and Mark Watson.
\newblock A buffer-based approach to rate adaptation: Evidence from a large
  video streaming service.
\newblock In {\em Proc. ACM SIGCOMM}, 2014.

\bibitem{huang2012confused}
Te-Yuan Huang, Nikhil Handigol, Brandon Heller, Nick McKeown, and Ramesh
  Johari.
\newblock Confused, timid, and unstable: picking a video streaming rate is
  hard.
\newblock In {\em Proceedings of the 2012 ACM conference on Internet
  measurement conference}, pages 225--238. ACM, 2012.

\bibitem{muller2012evaluation}
Christopher M{\"u}ller, Stefan Lederer, and Christian Timmerer.
\newblock An evaluation of dynamic adaptive streaming over http in vehicular
  environments.
\newblock In {\em Proceedings of the 4th Workshop on Mobile Video}, pages
  37--42. ACM, 2012.

\bibitem{riiser2012comparison}
Haakon Riiser, H{\aa}kon~S Bergsaker, Paul Vigmostad, P{\aa}l Halvorsen, and
  Carsten Griwodz.
\newblock A comparison of quality scheduling in commercial adaptive http
  streaming solutions on a 3g network.
\newblock In {\em Proceedings of the 4th Workshop on Mobile Video}, pages
  25--30. ACM, 2012.

\bibitem{nicholson2008breadcrumbs}
Anthony~J Nicholson and Brian~D Noble.
\newblock Breadcrumbs: forecasting mobile connectivity.
\newblock In {\em Proceedings of the 14th ACM international conference on
  Mobile computing and networking}, pages 46--57. ACM, 2008.

\bibitem{xu2013proteus}
Qiang Xu, Sanjeev Mehrotra, Zhuoqing Mao, and Jin Li.
\newblock Proteus: network performance forecast for real-time, interactive
  mobile applications.
\newblock In {\em Proceeding of the 11th annual international conference on
  Mobile systems, applications, and services}, pages 347--360. ACM, 2013.

\bibitem{balachandran2012quest}
Athula Balachandran, Vyas Sekar, Aditya Akella, Srinivasan Seshan, Ion Stoica,
  and Hui Zhang.
\newblock A quest for an internet video quality-of-experience metric.
\newblock In {\em Proceedings of the 11th ACM Workshop on Hot Topics in
  Networks}, pages 97--102. ACM, 2012.

\bibitem{ITUDoc}
Methodology for the subjective assessment of the quality of television
  pictures.
\newblock Recommendation ITU-R BT.500-13, 2012.

\bibitem{riiser2013commute}
Haakon Riiser, Paul Vigmostad, Carsten Griwodz, and P{\aa}l Halvorsen.
\newblock Commute path bandwidth traces from 3g networks: analysis and
  applications.
\newblock In {\em Proceedings of the 4th ACM Multimedia Systems Conference},
  pages 114--118. ACM, 2013.

\end{thebibliography}
\end{document}